%% file: paper.tex
\documentclass[12pt]{article}
\usepackage{epsfig,graphicx,latexsym,url,wrapfig}

\title{Ununfoldable Polyhedra with Convex Faces}

\author{
      Marshall Bern%
        \thanks{Xerox Palo Alto Research Center, 3333 Coyote Hill Rd.,
                Palo Alto, CA 94304, USA, email:
                \texttt{bern@parc.xerox.com}.}
 \and Erik D. Demaine%
        \thanks{Department of Computer Science, University of Waterloo,
                Waterloo, Ontario N2L 3G1, Canada, email:
                \texttt{eddemaine@\penalty \exhyphenpenalty uwaterloo.ca}.}
 \and David Eppstein%
        \thanks{Department of Information and Computer Science,
                University of California, Irvine, CA 92697, USA,
                email: \texttt{eppstein@\penalty \exhyphenpenalty ics.uci.edu}.}
 \andlinebreak Eric Kuo%
        \thanks{EECS Computer Science Division, University of California,
                Berkeley, 387 Soda Hall \#1776, Berkeley, CA 94720-1776, USA,
                email:
                \texttt{ekuo@\penalty \exhyphenpenalty eecs.berkeley.edu}.
                Work performed while at Xerox PARC.}
 \and Andrea Mantler%
        \thanks{Department of Computer Science, University of British Columbia,
                Vancouver, BC V6T 1Z4, Canada,
                email: \texttt{mantler@\penalty \exhyphenpenalty cs.unc.edu}.
                Work supported in part by NSERC and performed while visiting
                UNC.}
 \and Jack Snoeyink%
        \footnotemark[5]~
        \thanks{Department of Computer Science, University of North Carolina,
                Chapel Hill, NC 27599-3175, USA,
                email: \texttt{snoeyink@\penalty \exhyphenpenalty cs.unc.edu}.}
}

\date{}

\advance \topmargin by -\headheight
\advance \topmargin by -\headsep
\evensidemargin \oddsidemargin
\def\andlinebreak{\end{tabular}\linebreak\begin{tabular}[t]{c}}

{\makeatletter \gdef\fps@figure{htbp}}

\newtheorem{lemma}{Lemma}
\newtheorem{theorem}[lemma]{Theorem}
\newtheorem{corollary}[lemma]{Corollary}
\newenvironment{proof}{\textbf{Proof: }\ignorespaces}
  {\hspace*{\fill}$\Box$\medskip}

\begin{document}
\maketitle

\begin{abstract}
Unfolding a convex polyhedron into a simple planar polygon is a well-studied problem.
In this paper, we study the limits of unfoldability by
studying nonconvex polyhedra with the same combinatorial structure as convex
polyhedra.  In particular, we give two examples of polyhedra, one with 24
convex faces and one with 36 triangular faces, that cannot be
unfolded by cutting along edges.  We further show that such a polyhedron can
indeed be unfolded if cuts are allowed to cross faces.  Finally, we prove that
``open'' polyhedra with triangular faces may not be unfoldable no matter how
they are cut.
\end{abstract}

\section{Introduction}

A classic open question in geometry
\cite{Croft-Falconer-Guy-1991-unfolding,Fukuda-1997,O'Rourke-1998,Shephard-1975}
is whether every convex polyhedron can be cut along its edges and flattened
into the plane without any overlap.  Such a collection of cuts is called an
\emph{edge cutting} of the polyhedron, and the resulting simple polygon is
called an \emph{edge unfolding} or \emph{net}.  While the first explicit
description of this problem is by Shephard in 1975 \cite{Shephard-1975}, it has
been implicit since at least the time of Albrecht D\"urer, circa 1500
\cite{Duerer-1525}.

It is widely conjectured that every convex polyhedron has an edge unfolding.
Some recent support for this conjecture is that every triangulated convex
polyhedron has a \emph{vertex unfolding}, in which the cuts are along
edges but the unfolding only needs to be connected at vertices
\cite{Demaine-Eppstein-Erickson-Hart-O'Rourke-2001}.  On the other hand,
experimental results suggest that a random edge cutting of a random polytope
causes overlap with probability approaching~$1$ as the number of vertices
approaches infinity \cite{Schevon-1989}.
% page 30

While unfoldings were originally used to make paper models of polyhedra
\cite{Cundy-Rollett-1961-polyhedra, Wenninger-1971}, unfoldings have important
industrial applications.
For example, sheet metal bending is an efficient process for manufacturing
\cite{Gupta-Bourne-Kim-Krishnan-1998, Wang-1997}.  In this process, the desired
object is approximated by a polyhedron, which is unfolded into a collection of
polygons.  Then these polygons are cut out of a sheet of material, and each
piece is folded into a portion of the object's surface using a bending machine.
The unfoldings have multiple pieces partly for practical reasons such as
efficient packing into a rectangle of material, but mainly because little
theory on unfolding nonconvex polyhedra is available, and thus heuristics must
be used \cite{O'Rourke-1998}.

There are two freely available heuristic programs for constructing edge
unfoldings of polyhedra: the Mathematica package UnfoldPolytope
\cite{Namiki-Fukuda-1994}, and the Macintosh program HyperGami
\cite{Hayes-1995-hypergami}.  There are no reports of these programs failing to
find an edge unfolding for a convex polyhedron; HyperGami even finds unfoldings
for nonconvex polyhedra.  There are also several commercial heuristic
programs; an example is Touch-3D \cite{Lundstrom-Design-1998}, which supports
nonconvex polyhedra by using multiple pieces when needed.

It is known that if we allow cuts across the faces as well as along the edges,
then every \emph{convex} polyhedron has an unfolding.  Two such unfoldings are
known.  The simplest to describe is the \emph{star unfolding}
\cite{Agarwal-Aronov-O'Rourke-Schevon-1997,Aronov-O'Rourke-1992}, which cuts
from a generic point on the polyhedron along shortest paths to each of the
vertices.  The second is the \emph{source unfolding}
\cite{Mitchell-Mount-Papadimitriou-1987, Sharir-Schorr-1986}, which cuts along
points with more than one shortest path to a generic source point.

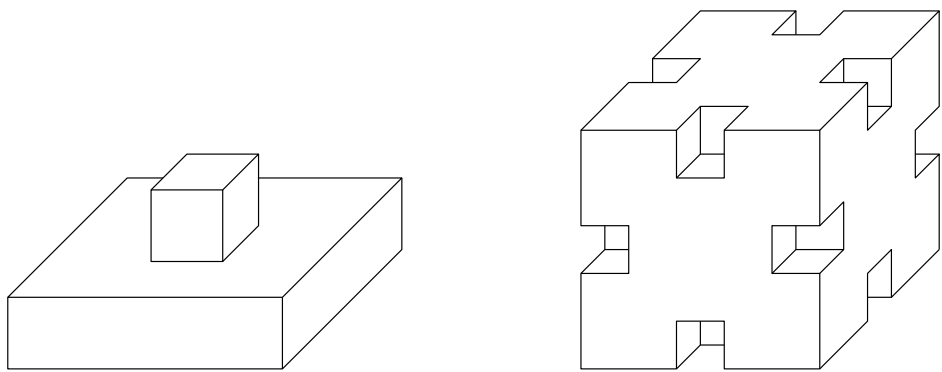
\begin{figure}
\centerline{\input{bad_orthos.pstex_t}}
\caption{\label{bad orthos} Orthogonal polyhedra with no edge unfoldings
  from \protect\cite{CCCG98b}.}
\end{figure}

There has been little theoretical work on unfolding nonconvex polyhedra.
In what may be the only paper on this subject,
Biedl et al.~\cite{CCCG98b} show the positive result that
certain classes of orthogonal polyhedra can be unfolded.  They show the
negative result that not all nonconvex polyhedra have edge
unfoldings.  Two of their examples are given in Figure~\ref{bad orthos}.  The
first example is rather trivial: the top box must unfold to fit inside the
hole of the top face of the bottom box, but there is insufficient area to do
so.  The second example is closer to a convex polyhedron in the sense that
every face is homeomorphic to a disk.

Neither of these examples is satisfying because they are not ``topologically
convex.''  
A polyhedron is \emph{topologically convex} if
its graph (1-skeleton)
is isomorphic to the graph of a convex polyhedron.
The first example of Figure~\ref{bad orthos}
is ruled out because faces of a convex polyhedron are always homeomorphic to
disks.  The second example is ruled out because there are pairs of faces that
share more than one edge, which is impossible for a convex
polyhedron\footnote{We disallow polyhedra with ``floating vertices,'' that is,
vertices with angle $\pi$ on both incident faces, which can just be removed.}.
In general, a famous theorem of Steinitz \cite{Gruenbaum-1967,
Henk-Richter-Gebert-Ziegler-1997, Steinitz-Rademacher-1976} tells us that a
polyhedron is topologically convex precisely if its graph is 3-connected and
planar.

The class of topologically convex polyhedra includes all \emph{convex-faced}
polyhedra (i.e., polyhedra whose faces are all convex) that are homeomorphic to
spheres.  Schevon and other researchers \cite{CCCG98b,Schevon-1987} have asked
whether all such polyhedra can be unfolded without overlap by cutting along
edges.  In other words, can the conjecture that every convex polyhedron is
edge-unfoldable be extended to topologically convex polyhedra?  Another
particularly interesting subclass, which we consider here, are polyhedra whose
faces are all triangles (called \emph{triangulated} or \emph{simplicial}) and
are homeomorphic to spheres.

In this paper we construct families of triangulated and convex-faced polyhedra
that are homeomorphic to spheres and have no edge unfoldings.  This proves, in
particular, that the edge-unfolding conjecture does not generalize to
topologically convex polyhedra.  We go on to show that cuts across faces can
unfold some convex-faced polyhedra that cannot be unfolded with cuts only along
edges.  This is the first demonstration
that general cuts are more powerful than edge cuts.

We also consider the problem of constructing a polyhedron that cannot be
unfolded even using general cuts.  If such a polyhedron exists, the theorem
that every convex polyhedron is generally unfoldable (using, for example, the
star or source unfolding) cannot be extended to topologically convex polyhedra.
As a step towards this goal, we present an ``open'' triangulated polyhedron
that cannot be unfolded.  Finding a ``closed'' ununfoldable polyhedron is an
intriguing open problem.

A preliminary version of this work appeared in CCCG'99
\cite{Bern-Demaine-Eppstein-Kuo-1999-both}.

\section{Basics}
\label{Basics}

We begin with formal definitions and some basic results about polyhedra,
unfoldings, and cuttings.

We define a \emph{polyhedron} to be a
connected set of closed planar polygons in 3-space such that
(1) any intersection between two polygons in the set
    is a collection of vertices and edges common to both polygons, and
(2) each edge is shared by at most two polygons in the set.
A \emph{closed} polyhedron 
is one in which each edge is shared by exactly two
polygons.  If a polyhedron is not closed,
we call it an \emph{open} polyhedron;
the \emph{boundary} of an open polyhedron is the set of edges
covered by only one polygon.

In general, we may allow the polygonal faces to be multiply connected (i.e.,
have holes).  However, in this paper we concentrate on convex-faced polyhedra.
A polyhedron is \emph{convex-faced} if every face is strictly convex, that is,
every interior face angle is strictly less than $\pi$.  Thus, in
particular, every face of a convex-faced polyhedron is simply connected.  A
polyhedron is \emph{triangulated} if every face is a triangle, that is, has
three vertices.

Next we prove that convex-faced polyhedra (and hence triangulated polyhedra)
are a subclass of \emph{topologically convex} polyhedra, that is, polyhedra
whose graphs (1-skeleta) are isomorphic to graphs of convex polyhedra.
A \emph{convex polyhedron} is a closed polyhedron whose interior is a convex
set---equivalently, the open line segment connecting any pair of points on the
polyhedron's surface is interior to the polyhedron.

\begin{theorem}[Steinitz's Theorem \rm \cite{Gruenbaum-1967,
Henk-Richter-Gebert-Ziegler-1997, Steinitz-Rademacher-1976}]
%--- Surprisingly, \cite{Cromwell-1997} does not cover this theorem.  -Erik
A graph is the graph of a convex polyhedron precisely if it is 3-connected and
planar.
\end{theorem}

\begin{corollary}
Every convex-faced closed polyhedron that is homeomorphic to a sphere is
topologically convex.
\end{corollary}

\begin{proof}
Because the polyhedron is homeomorphic to a sphere, its graph $G$ must be
planar.  It remains to show that $G$ is 3-connected.  If there is a vertex $v$
whose removal disconnects $G$ into at least two components $G_1$ and $G_2$,
then there must be a ``belt'' wrapping around the polyhedron that separates
$G_1$ and $G_2$.  This belt has only one vertex ($v$) connecting $G_1$ and
$G_2$, so it must consist of a single face.  This face touches itself at $v$,
which is impossible for a convex polygon.  Similarly, if there is a pair $(v_1,
v_2)$ of vertices whose removal disconnects $G$ into at least two components
$G_1$ and $G_2$, then again there must be a ``belt'' separating the two
components, but this time the belt connects $G_1$ and $G_2$ at two vertices
($v_1$ and $v_2$).  Thus, the belt can consist of up to two faces, and these
faces share two vertices $v_1$ and $v_2$.  This is impossible for strictly
convex polygons.
\end{proof}

A \emph{cutting} of a polyhedron $P$ is a union
$C$ of a finite number of line segments (called \emph{cuts}) on $P$,
such that cutting along $C$ results in a connected surface $P-C$
that can be flattened into the plane (that is, isometrically
embedded) without overlap.  The resulting flattened form is called
an \emph{unfolding} of $P$.
An \emph{edge cutting} of $P$ is a cutting of $P$
that is just a union of $P$'s edges.
The corresponding unfolding is called an \emph{edge unfolding}.
We sometimes call unfoldings \emph{general unfoldings}
to distinguish them from edge unfoldings.
We call a polyhedron 
\emph{unfoldable} if it has a general unfolding, and \emph{edge-unfoldable}
if it has an edge unfolding.
Similarly, we call a polyhedron [\emph{edge}-]\emph{ununfoldable}
if it is not [edge-]unfoldable.

If $P$ is an open polyhedron, we limit attention to cuttings
that contain only a finite number of boundary points of $P$.
Every neighborhood of a boundary point of $P$
in $C$ must contain an interior point of $P$ in $C$; otherwise,
the boundary point could be removed from $C$
without changing $P-C$.
Hence, the collection of boundary points of $P$ in $C$ can be reduced down
to a finite set without any effect.

We define the \emph{curvature} of an interior vertex $v$ to be the discrete
analog of Gaussian curvature, namely $2 \pi$ minus the sum of the face angles
at $v$.  (For vertices on the boundary of the polyhedron, we do not define
the notion of curvature.)
% we count the incident face angles on both sides.)
Hence, the neighborhood of a zero-curvature vertex
can be flattened into the plane, the neighborhood of a positive-curvature
vertex (for example, a cone) requires a cut in order to be flattened, and the
neighborhood of a negative-curvature vertex (for example, a saddle) requires
two or more cuts to avoid self-overlap.

We look at a cutting as a graph drawn on the polyhedron, and establish some
basic facts about these graphs in the next lemmas.

\begin{lemma} \label{spanning forest}
Any cutting of a polyhedron $P$ is a forest, and spans every nonboundary
vertex of $P$ that has nonzero curvature. 
\end{lemma}

\begin{proof}
If a cutting contained a cycle, having positive area both interior and
exterior to the cycle, then the resulting unfolding would be disconnected,
a contradiction.
(Above we excluded the possibility of the cutting containing a cycle on the
boundary of an open polyhedron $P$.)
If the cutting did not contain a particular (nonboundary)
point with nonzero curvature, neighborhoods of that point could not be
flattened without overlap.
\end{proof}

\begin{figure}
\centerline{\includegraphics{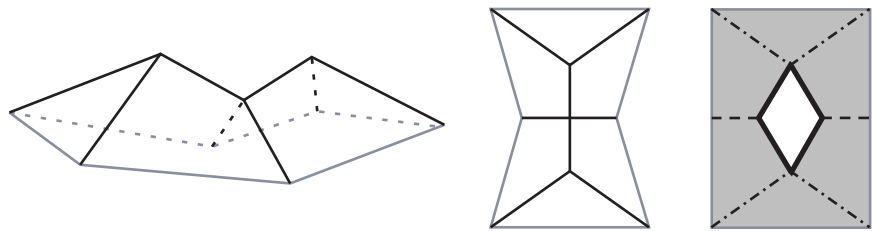}}
\caption{\label{forestcut}
  A portion of a polyhedron in (left) perspective view and (middle)
  birds-eye view, and (right) the unfolding that results from a cutting
  with a separated connected component of cuts.  To formally construct this
  portion of a polyhedron, start with the unfolding on the right,
  and fold as shown.}
\end{figure}

A common assumption is that an unfolding must be a simple polygon,
which implies that a cutting of a closed polyhedron must furthermore be a
(connected) tree; see e.g.\ \cite{Demaine-Demaine-Lubiw-O'Rourke-2000-TR,
Demaine-Demaine-Lubiw-O'Rourke-2000, Schevon-1989}.
%\cite[Lemma~2.2.1]{Schevon-1989}.
Without this restriction, in most cases, a cutting of a closed polyhedron
is a tree, but in fact this is not always the case.
Figure~\ref{forestcut} illustrates a basic construction for
separating off a connected component of the cutting.  We could add cuts to
connect the inner tree to the rest of the forest, but these extra cuts are
unnecessary.  Using this construction, we can build a polyhedron and a
cutting having arbitrarily many connected components, by connecting a series
of the constructions in Figure~\ref{forestcut} in a ``dented barrel'' shape,
and then capping the ends.

Together with Joseph O'Rourke (personal communication, July 2001),
we have proved that such examples require nonconvexity:

\begin{lemma}
Any cutting of a convex closed polyhedron is a tree.
\end{lemma}

\begin{proof}
By Lemma \ref{spanning forest}, the cutting is a forest.
Suppose for contradiction that the forest has multiple connected components.
Then there is a closed path $p$ on the surface of the polyhedron that
avoids all cuts and strictly encloses a connected component of the cutting.
In particular, $p$ avoids all vertices of the polyhedron.
Let $\tau$ denote the total turn angle along the path $p$.
Because $p$ avoids vertices, it unfolds to a connected (uncut) closed path
in the planar layout; thus, $\tau = 2 \pi$.
Because the polyhedron is homeomorphic to a sphere,
the Gauss-Bonnet theorem \cite[pp.~215--217]{Cromwell-1997}
applied to $p$ says that
$\tau + \gamma = 2\pi$ where $\gamma$ is the curvature enclosed by $p$.
By convexity, $\gamma \geq 0$.  Further, $p$ encloses a connected
component of the cutting, which consists of at least one vertex of the
polyhedron, so $\gamma > 0$.  Therefore, $\tau < 2\pi$, contradicting that
$p$ could lay out flat in the plane.
\end{proof}

\begin{lemma} \label{negative curvature}
If $v$ is a vertex of a polyhedron $P$ with negative curvature, then any
cutting of $P$ must include more than one cut incident to $v$.
\end{lemma}

\begin{proof}
Suppose some cutting $C$ includes only a single cut
incident to $v$.  Let $N$ be $P \cap B$, where $B$ is a small
ball around $v$.  Neighborhood $N$ unfolds to a small disk
that self-overlaps by precisely the absolute value of the curvature of $v$.
\end{proof}

\section{Hats}

Our construction of a closed polyhedron with no edge unfolding begins by
constructing open polyhedra that cannot be edge unfolded.  The intuition is
to build a
``hat-shaped''
polyhedron
%--- Erik recalls the drive from Marshall's house to Xerox PARC when Marshall
%    suggested this idea, which is the basis for all our examples!
having just one interior vertex with positive curvature (the peak of the hat).
The remaining vertices have negative curvature and this severely limits
the possible edge cuttings.

\begin{figure}
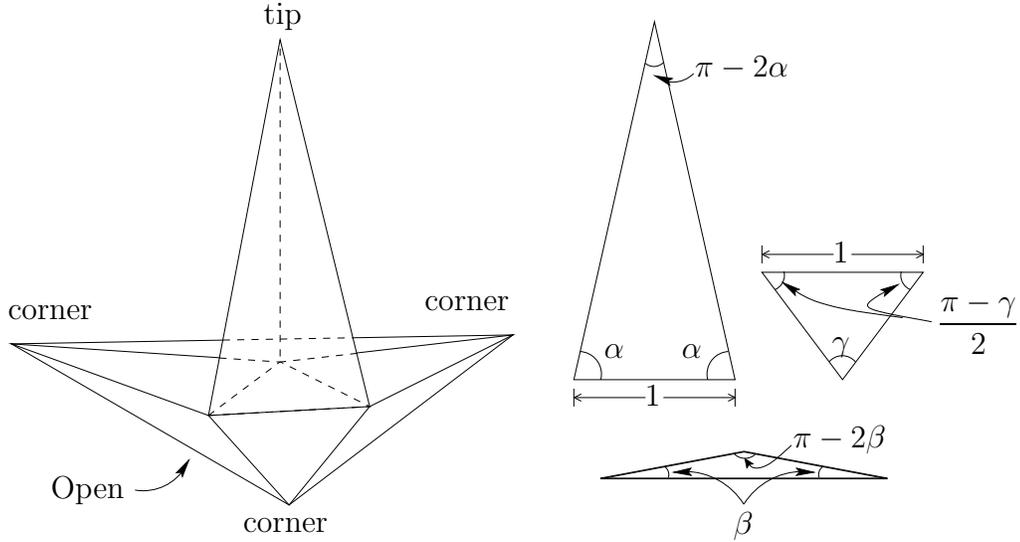

\centerline{\input trihat_details.pstex_t}
\caption{\label{triangulated hat}
  A triangulated hat and its constituent faces.}
\end{figure}

We know of two combinatorially different families of convex-faced \emph{hats}
that suffice to prevent edge cuttings.  They are shown in Figures~\ref{basic
hat} and~\ref{triangulated hat}.  The remainder of this section defines and
analyzes hats in detail.  These hats have the additional property that their
boundary is a triangle, and in the next section we will exploit this by gluing
each hat to a face of a regular tetrahedron.

\begin{figure}
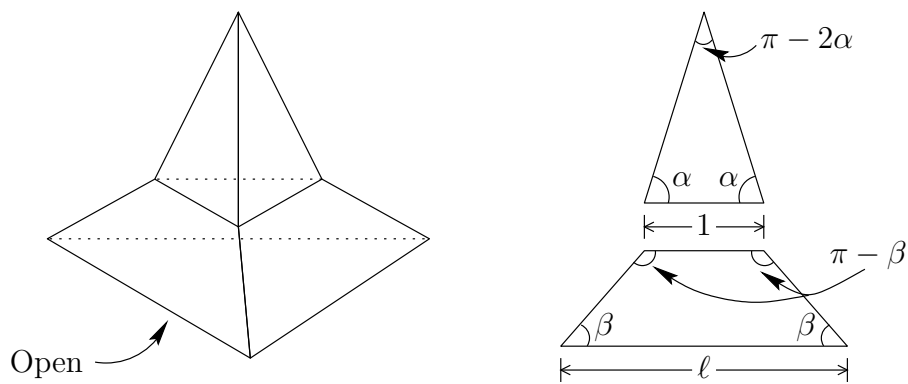

\centerline{\input hat_details.pstex_t}
\caption{\label{basic hat}
  A basic hat and its constituent faces.}
\end{figure}

The first hat (Figure~\ref{basic hat}) is called a \emph{basic hat} because
of its low face count.  The second hat (Figure~\ref{triangulated hat}),
however, has only triangular faces, and is hence called a \emph{triangulated
hat}.  In both cases, we will eventually find a convex-faced closed polyhedra
with no edge unfoldings, and in the latter case it will also be
triangulated.

The two types of hats are similar, so we will treat them together.  First we
need some terminology for features of the hats.  The three 
boundary
vertices are called the \emph{corners} of the hat, the innermost vertex is the 
\emph{tip}, and the 
remaining vertices
are the \emph{middle} vertices.  The three triangles incident to
the tip form the \emph{spike} of the hat.  The remaining faces (incident to the
corners) form the \emph{brim} of the hat.  The brim of a basic hat consists of
three trapezoids, whereas the brim of a triangulated hat consists of six
triangles.  Note that the spike of the triangulated hat is rotated $60^\circ$
relative to the boundary, in contrast to the basic hat.

More formally, hats are parameterized by three parameters.  Basic hats are
parameterized by angles $\alpha$ and $\beta$ satisfying $30^\circ \leq
\alpha,\beta < 90^\circ$, and a length $\ell > 1$.  For consistency with the
other type of hat, we also define $\gamma=0$ for basic hats.  The spike is an
open tetrahedron consisting of three identical isosceles triangles, where the
lower angles are $\alpha$ and the bottom sides have unit length.  The brim is
an open truncated tetrahedron consisting of three identical symmetric
trapezoids, where the lower angles are $\beta$, the top length is 1, and the
bottom length is $\ell$.

Triangulated hats are parameterized by angles $\alpha$, $\beta$, and $\gamma$
satisfying $30^\circ \leq \alpha, \beta+\gamma/2 < 90^\circ$ and $\gamma <
60^\circ$.  The spike is an open tetrahedron with base angles $\alpha$, just
like the basic hat.  In the brim, there are three identical isosceles triangles
touching the boundary along their base edges, which have base angles $\beta$;
and three identical isosceles triangles touching the peak at their base edges,
and touching the boundary at their opposite vertex, whose angle is $\gamma$.

As mentioned above, the key to our constructions is to make the middle vertices
have negative curvature.

\begin{lemma} \label{middle vertices have negative curvature}
The middle vertices of a hat have negative curvature precisely if $\alpha >
\beta + \gamma/2$.  In particular, for any valid choice of $\beta$ and
$\gamma$, there is a valid choice of $\alpha$ that satisfies this property.
\end{lemma}

\begin{proof}
The first 
claim
follows simply by summing the angles incident to a middle
vertex and checking when that sum is greater than $2 \pi$.  For basic hats, we
have
  $$2 \alpha + 2 (\pi - \beta) > 2 \pi$$
and for triangulated hats, we have
  $$2 \alpha + 2 \frac{\pi-\gamma}{2} + \pi - 2\beta > 2 \pi.$$

Now $\alpha$ is only restricted by $30^\circ \leq \alpha < 90^\circ$ for
validity and $\alpha > \beta+\gamma/2$ for negative curvature.  Because
$30^\circ \leq \beta+\gamma/2$, these two conditions are satisfied precisely if
$\beta+\gamma/2 < \alpha < 90^\circ$, which is satisfiable because
$\beta+\gamma/2 < 90^\circ$.
\end{proof}

Note that these conditions are rather symmetric for the two types of hats,
specifying that the angle at a base vertex of the spike is larger than the
angle at a corner of the hat.

A more useful property of hats is that the middle vertices can be made to have
negative curvature even when one of the spike triangles is cut away:

\begin{lemma} \label{stronger negative curvature}
The middle vertices of a hat have negative curvature even when a spike
triangle is removed, provided $\alpha > 2 \beta + \gamma$.  In particular, for
any $\beta$ and $\gamma$ with $30^\circ \leq \beta + \gamma/2 < 45^\circ$ and
$\gamma < 60^\circ$, there is a valid choice of $\alpha$ that satisfies these
properties.
\end{lemma}

\begin{proof}
The first 
claim
follows from Lemma~\ref{middle vertices have negative
curvature} by halving the coefficient of $\alpha$, because only a single
$\alpha$ is now included in the sum of angles.  The constraints now become
$30^\circ \leq \alpha < 90^\circ$ and $\alpha > 2 \beta + \gamma$.  Because
$30^\circ \leq \beta + \gamma/2$, these are equivalent to $2 \beta + \gamma <
\alpha < 90^\circ$, which is achievable because $\beta+\gamma/2 < 45^\circ$.
\end{proof}

For example, a hat with angles $\alpha = 81^\circ$, $\beta = 30^\circ$,
and 
$\gamma = 20^\circ$, has negative curvature at the middle vertices even when a 
spike triangle is removed.  In the case of the basic hat, we would have 
$\alpha + 2(\pi - \beta) 
%(AM) = 81^\circ + 2(180^\circ - 30^\circ) 
= 381^\circ$, 
and for the triangulated hat $\alpha + (\pi - 2\beta) + 2(\pi - \gamma)/2 
%(AM) = 81^\circ + (180^\circ - 2(30^\circ)) + (180^\circ - 20^\circ) 
= 361^\circ$.

\begin{theorem} \label{open edge-ununfoldable}
Hats that satisfy the constraints in Lemma~\ref{stronger negative curvature} 
are open convex-faced polyhedra with no edge unfoldings.
\end{theorem}

\begin{figure}
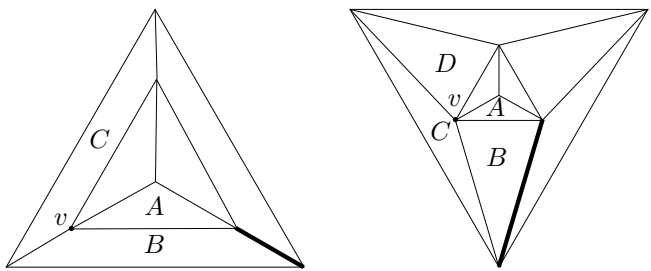

\centerline{\input planar_proof.pstex_t}
\caption{\label{hat proof}
  Proof of Lemma~\ref{open edge-ununfoldable}:
  a planar drawing of each kind of hat with a possible cut in bold.}
\end{figure}

\begin{proof}
By Lemma~\ref{spanning forest}, any edge cutting is a forest of nonboundary
edges that covers the tip and middle vertices.  Every connected component of
the cutting is a tree, and so must have at least two leaves.  Note that no
two corners of the hat can be leaves of a common connected component of the
cutting, because otherwise the path of cuts connecting them would disconnect
the polyhedron.  (Recall we excluded the possibility of a boundary edge being
in a cutting.)  Thus, at most one corner is a leaf of each connected
component of the cutting.  By Lemma~\ref{middle vertices have negative
curvature}, the middle vertices have negative curvature, so by 
Lemma~\ref{negative curvature}, they cannot be leaves of the cutting.  Hence, 
the cutting must in fact be a single path from a corner to the tip, visiting 
all of the middle vertices.

It is possible to argue by case analysis that, for basic hats, there is
precisely one such path up to symmetry (see Figure~\ref{spiral unfolding},
left), and for triangulated hats, there are two such paths up to symmetry (see
Figure~\ref{spiral unfolding}, center and right).  
Each
of these cuttings has a vertex with only one spike triangle cut away (marked 
by a gray circle in Figure~\ref{spiral unfolding}), which means the remainder 
has negative curvature, leading to overlap by Lemma~\ref{negative curvature}.

\begin{figure}
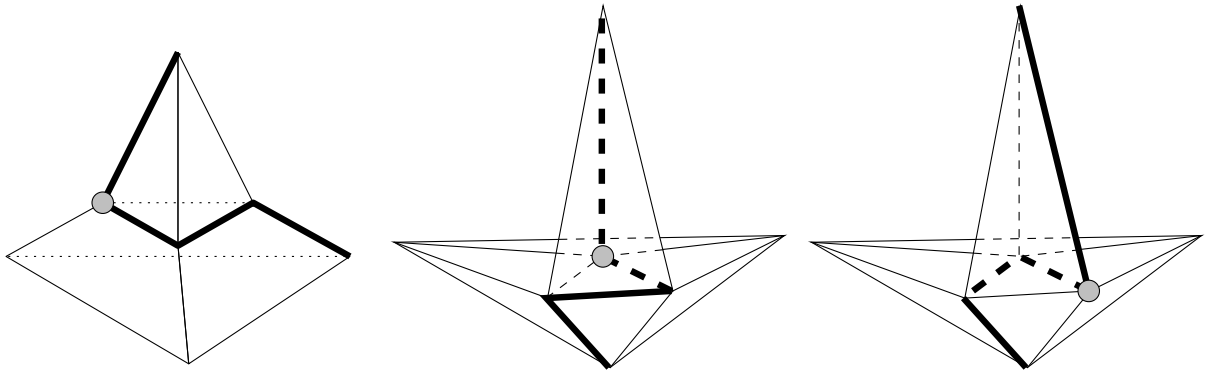

\centerline{\input spiral_cuts.pstex_t}
\caption{\label{spiral unfolding} The possible cuttings (up to symmetry)
  of a hat satisfying Lemma~\protect\ref{stronger negative curvature}.}
\end{figure}

However, this can be argued more simply as follows.  Because the spike remains
connected to the rest of the polyhedron, there must be a spike triangle $A$
that remains connected to a brim face $B$; see Figure~\ref{hat proof}.  
Because there is only one cut in the hat incident to a corner, this cut is not 
incident to one of the two vertices shared by faces $A$ and $B$, say $v$.
Therefore the brim faces $B$, $C$, and (in the case of a triangulated hat) $D$ 
incident to $v$ and the spike face $A$ remain connected in the unfolding along 
edges incident to $v$.  But by Lemma~\ref{stronger negative curvature}, these 
faces have total angle at $v$ of more than $360^\circ$, 
causing overlap.
\end{proof}

\section{Gluing Hats Together}

Because the boundary of a hat is an equilateral triangle,
we can take four hats and
place them against the faces of a regular tetrahedron (and then remove the
guiding tetrahedron).  The result
is a closed polyhedron with no edge unfolding,
which we call a \emph{spiked tetrahedron}.
First observe the following property of unfolded hats:

\begin{figure}
\centerline{%
  $\vcenter{\hbox{%\reflectbox{
                  \includegraphics[scale=0.75]{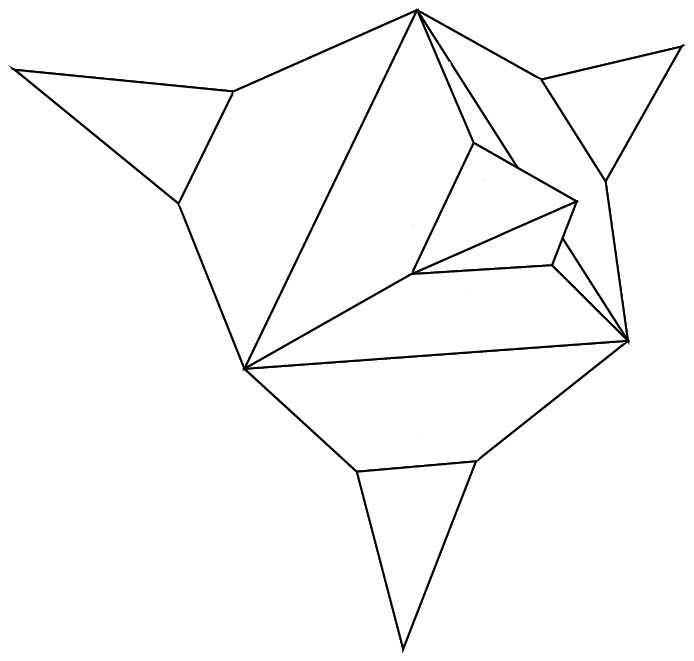}}%
                  %}
                 }$%
  \hfil
  $\vcenter{\hbox{\includegraphics[scale=0.8]{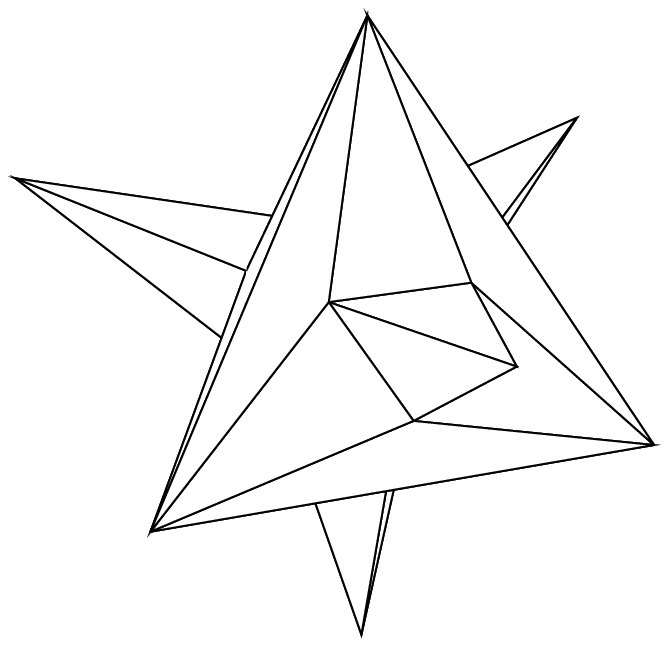}}}$%
}
\caption{\label{spiked tetrahedra} Spiked tetrahedra for both types of hats.}
\end{figure}

\begin{lemma} \label{corner-to-corner path}
In any edge cutting of a spiked tetrahedron, there is a path joining at least
two corners of each hat using nonboundary edges of that hat, provided the
parameters satisfy the constraint in Lemma~\ref{stronger negative curvature}.
\end{lemma}

\begin{proof}
This lemma follows directly from Theorem~\ref{open edge-ununfoldable}:
an edge cutting of the spiked tetrahedron cannot induce an edge cutting of
a constituent hat, and cannot cause overlap, so it must cut each hat into
at least two pieces, by way of a path with the claimed properties.
\end{proof}

Now the desired result follows easily:

\begin{theorem} \label{closed edge-ununfoldable}
Spiked tetrahedra are convex-faced closed polyhedra with no edge unfoldings,
provided the constituent hats satisfy the constraint in Lemma~\ref{stronger
negative curvature}.
\end{theorem}

\begin{proof}
Suppose there were an edge cutting.  By Lemma~\ref{corner-to-corner path},
inside each of the four hats would be paths of cuts joining two corners, and
these paths share no cuts because they use only nonboundary edges.  But because
there are only four corners in total, these paths would form a cycle in the
cutting, contradicting Lemma~\ref{spanning forest}.
\end{proof}

This theorem also proves that the analogously defined \emph{spiked octahedron}
cannot be edge unfolded for hats satisfying the constrain in Lemma~\ref{stronger
negative curvature}.  In particular, there is a basic-hat spiked octahedron
having no edge unfolding, for any $\beta$ satisfying $30^\circ \leq \beta <
45^\circ$.  Using a different proof technique
\cite{Bern-Demaine-Eppstein-Kuo-1999-both}, it can be shown that there is such
a spiked octahedron for any $\beta$ satisfying $45^\circ < \beta < 60^\circ$.

\section{General Unfolding}
\label{General Unfolding}

While spiked tetrahedra are not edge-unfoldable for certain choices of the
parameters, they are always generally unfoldable; see Figure~\ref{general
unfolding}.  We start with the parallelogram unfolding of
the underlying tetrahedron on which we glued the hats.
The spikes do not have room to unfold in the middle of an unfolded brim, so we
use the following trick.  Cut out each spike and a small band that connects it
to an edge on the boundary of the tetrahedron unfolding.  Now attach that band
and unfolded spike to the corresponding edge of the tetrahedron unfolding
(corresponding in the sense that the edges are glued to each other).  The bands
are chosen to be nonperpendicular so that a band does not attempt to attach
where another band was removed.

\begin{figure}
\centerline{\includegraphics[scale=0.5]{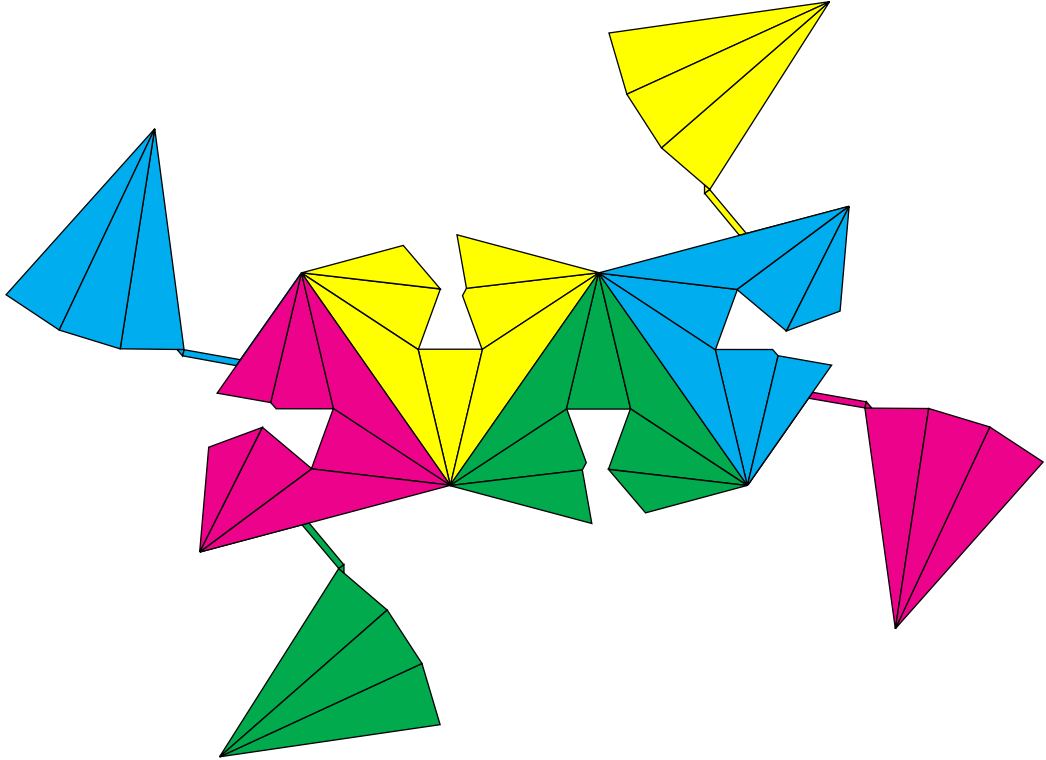}}
\caption{\label{general unfolding} General unfolding of a spiked tetrahedron.}
\end{figure}

\section{No General Unfolding}
\label{No General Unfolding}

An intriguing open question is whether there is a convex-faced polyhedron,
triangulated polyhedron, or any closed polyhedron that cannot be generally
unfolded.  This section makes a step toward solving this problem by presenting
a triangulated \emph{open} polyhedron (polyhedron with boundary) with no
general unfolding.

\begin{figure}
\centerline{
	\begin{picture}(413,128)
		\put(0,10){\input easy_open.pstex_t}
		\put(185,0){\includegraphics{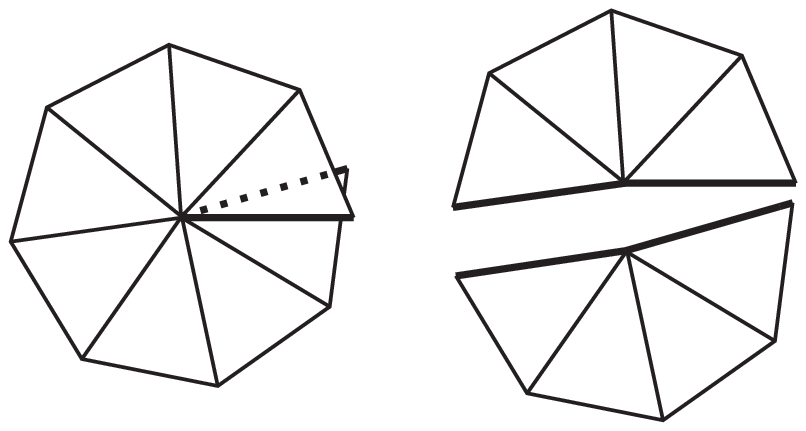}}
	\end{picture}
}
\caption{\label{easy open ununfoldable}
  Open polyhedron with no unfolding.
  One cut creates an unfolding with overlap, 
  but two cuts disconnect the results.}
\end{figure}

The construction is to connect several triangles in a cycle, all sharing a
common vertex $v$, as shown in Figure~\ref{easy open ununfoldable}.  By
connecting enough triangles and/or adjusting the triangles to have large enough
angle incident to $v$, we can arrange for vertex $v$ to have negative
curvature.

\begin{theorem} \label{open generally ununfoldable}
The open polyhedron in Figure~\ref{easy open ununfoldable} has no general
unfolding if $v$ has negative curvature.
\end{theorem}

\begin{proof}
A cutting could only have leaves on the boundary, because $v$ has negative
curvature, and because the cut incident to any other leaf could be glued
(uncut) without affecting the unfolding.  But any cutting has at least two
leaves, so it must disconnect the polyhedron, a contradiction.
\end{proof}

\section{Conclusion}

The spiked tetrahedra (Figure~\ref{spiked tetrahedra}) show that the conjecture
about edge unfoldings of convex polyhedra cannot be extended to topologically
convex polyhedra.  Figure~\ref{general unfolding} further illustrates the added
power of cuts along faces
for topologically convex polyhedra.  Figure~\ref{easy open ununfoldable} shows
an open polyhedron that cannot be unfolded at all, but the flexibility
exploited in Figure~\ref{general unfolding} suggests that it will be more
difficult to settle whether there is a closed polyhedron with that property.
Another interesting open question is the complexity of deciding whether a given
triangulated polyhedron has an edge unfolding, now that we know that the answer
is not always ``yes.''

\section*{Acknowledgments}

We thank Anna Lubiw for helpful discussions.
We thank Joseph O'Rourke for his contributions to Section \ref{Basics}
and for many helpful comments on the paper.

\bibliography{polytopes,unfolding}
\bibliographystyle{plain}

\end{document}

%% file: bad_orthos.pstex_t
\begin{picture}(0,0)%
\epsfig{file=bad_orthos.pstex}%
\end{picture}%
\setlength{\unitlength}{3947sp}%
\begingroup\makeatletter\ifx\SetFigFont\undefined%
\gdef\SetFigFont#1#2#3#4#5{%
  \reset@font\fontsize{#1}{#2pt}%
  \fontfamily{#3}\fontseries{#4}\fontshape{#5}%
  \selectfont}%
\fi\endgroup%
\begin{picture}(5874,2274)(2689,-2698)
\end{picture}

%% file: hat_details.pstex_t
\begin{picture}(0,0)%
\epsfig{file=hat_details.pstex}%
\end{picture}%
\setlength{\unitlength}{3947sp}%
\begingroup\makeatletter\ifx\SetFigFont\undefined%
\gdef\SetFigFont#1#2#3#4#5{%
  \reset@font\fontsize{#1}{#2pt}%
  \fontfamily{#3}\fontseries{#4}\fontshape{#5}%
  \selectfont}%
\fi\endgroup%
\begin{picture}(5338,2349)(3185,-4798)
\put(3601,-4711){\makebox(0,0)[rb]{\smash{\SetFigFont{12}{14.4}{\familydefault}{\mddefault}{\updefault}Open}}}
\put(7501,-3848){\makebox(0,0)[b]{\smash{\SetFigFont{12}{14.4}{\familydefault}{\mddefault}{\updefault}1}}}
\put(7720,-3544){\makebox(0,0)[rb]{\smash{\SetFigFont{12}{14.4}{\familydefault}{\mddefault}{\updefault}$\alpha$}}}
\put(7300,-3544){\makebox(0,0)[lb]{\smash{\SetFigFont{12}{14.4}{\familydefault}{\mddefault}{\updefault}$\alpha$}}}
\put(7856,-2674){\makebox(0,0)[lb]{\smash{\SetFigFont{12}{14.4}{\familydefault}{\mddefault}{\updefault}$\pi-2\alpha$}}}
\put(6813,-4479){\makebox(0,0)[lb]{\smash{\SetFigFont{12}{14.4}{\familydefault}{\mddefault}{\updefault}$\beta$}}}
\put(8204,-4479){\makebox(0,0)[rb]{\smash{\SetFigFont{12}{14.4}{\familydefault}{\mddefault}{\updefault}$\beta$}}}
\put(8290,-4021){\makebox(0,0)[lb]{\smash{\SetFigFont{12}{14.4}{\familydefault}{\mddefault}{\updefault}$\pi-\beta$}}}
\put(7501,-4748){\makebox(0,0)[b]{\smash{\SetFigFont{12}{14.4}{\familydefault}{\mddefault}{\updefault}$\ell$}}}
\end{picture}

%% file: trihat_details.pstex_t
\begin{picture}(0,0)%
\epsfig{file=trihat_details.pstex}%
\end{picture}%
\setlength{\unitlength}{2960sp}%
\begingroup\makeatletter\ifx\SetFigFont\undefined%
\gdef\SetFigFont#1#2#3#4#5{%
  \reset@font\fontsize{#1}{#2pt}%
  \fontfamily{#3}\fontseries{#4}\fontshape{#5}%
  \selectfont}%
\fi\endgroup%
\begin{picture}(7785,4419)(1126,-4186)
\put(1126,-2386){\makebox(0,0)[lb]{\smash{\SetFigFont{12}{14.4}{\rmdefault}{\mddefault}{\updefault}corner}}}
\put(2101,-3886){\makebox(0,0)[rb]{\smash{\SetFigFont{12}{14.4}{\familydefault}{\mddefault}{\updefault}Open}}}
\put(5326,-2311){\makebox(0,0)[rb]{\smash{\SetFigFont{12}{14.4}{\rmdefault}{\mddefault}{\updefault}corner}}}
\put(3425, 89){\makebox(0,0)[b]{\smash{\SetFigFont{12}{14.4}{\rmdefault}{\mddefault}{\updefault}tip}}}
\put(8101,-2648){\makebox(0,0)[b]{\smash{\SetFigFont{12}{14.4}{\familydefault}{\mddefault}{\updefault}$\gamma$}}}
\put(7701,-3458){\makebox(0,0)[lb]{\smash{\SetFigFont{12}{14.4}{\familydefault}{\mddefault}{\updefault}$\pi-2\beta$}}}
\put(6881,-349){\makebox(0,0)[lb]{\smash{\SetFigFont{12}{14.4}{\familydefault}{\mddefault}{\updefault}$\pi-2\alpha$}}}
\put(3451,-4171){\makebox(0,0)[b]{\smash{\SetFigFont{12}{14.4}{\rmdefault}{\mddefault}{\updefault}corner}}}
\put(8101,-1912){\makebox(0,0)[b]{\smash{\SetFigFont{12}{14.4}{\familydefault}{\mddefault}{\updefault}1}}}
\put(6526,-3112){\makebox(0,0)[b]{\smash{\SetFigFont{12}{14.4}{\familydefault}{\mddefault}{\updefault}1}}}
\put(7285,-4186){\makebox(0,0)[b]{\smash{\SetFigFont{12}{14.4}{\familydefault}{\mddefault}{\updefault}$\beta$}}}
\put(8911,-2493){\makebox(0,0)[lb]{\smash{\SetFigFont{12}{14.4}{\familydefault}{\mddefault}{\updefault}$\displaystyle{\pi-\gamma \over 2}$}}}
\put(6936,-2710){\makebox(0,0)[rb]{\smash{\SetFigFont{12}{14.4}{\familydefault}{\mddefault}{\updefault}$\alpha$}}}
\put(6117,-2710){\makebox(0,0)[lb]{\smash{\SetFigFont{12}{14.4}{\familydefault}{\mddefault}{\updefault}$\alpha$}}}
\end{picture}

%% file: spiral_cuts.pstex_t
\begin{picture}(0,0)%
\epsfig{file=spiral_cuts.pstex}%
\end{picture}%
\setlength{\unitlength}{3552sp}%
\begingroup\makeatletter\ifx\SetFigFont\undefined%
\gdef\SetFigFont#1#2#3#4#5{%
  \reset@font\fontsize{#1}{#2pt}%
  \fontfamily{#3}\fontseries{#4}\fontshape{#5}%
  \selectfont}%
\fi\endgroup%
\begin{picture}(8367,2615)(964,-2606)
\end{picture}

%% file: planar_proof.pstex_t
\begin{picture}(0,0)%
\epsfig{file=planar_proof.pstex}%
\end{picture}%
\setlength{\unitlength}{2368sp}%
\begingroup\makeatletter\ifx\SetFigFont\undefined%
\gdef\SetFigFont#1#2#3#4#5{%
  \reset@font\fontsize{#1}{#2pt}%
  \fontfamily{#3}\fontseries{#4}\fontshape{#5}%
  \selectfont}%
\fi\endgroup%
\begin{picture}(6742,2758)(1189,-3405)
\put(6360,-2296){\makebox(0,0)[b]{\smash{\SetFigFont{10}{12.0}{\rmdefault}{\mddefault}{\updefault}$B$}}}
\put(5827,-1309){\makebox(0,0)[b]{\smash{\SetFigFont{10}{12.0}{\rmdefault}{\mddefault}{\updefault}$D$}}}
\put(5777,-2019){\makebox(0,0)[b]{\smash{\SetFigFont{10}{12.0}{\rmdefault}{\mddefault}{\updefault}$C$}}}
\put(6346,-1769){\makebox(0,0)[b]{\smash{\SetFigFont{10}{12.0}{\rmdefault}{\mddefault}{\updefault}$A$}}}
\put(2776,-3211){\makebox(0,0)[b]{\smash{\SetFigFont{10}{12.0}{\rmdefault}{\mddefault}{\updefault}$B$}}}
\put(2776,-2806){\makebox(0,0)[b]{\smash{\SetFigFont{10}{12.0}{\rmdefault}{\mddefault}{\updefault}$A$}}}
\put(2206,-2101){\makebox(0,0)[b]{\smash{\SetFigFont{10}{12.0}{\rmdefault}{\mddefault}{\updefault}$C$}}}
\put(1801,-2911){\makebox(0,0)[b]{\smash{\SetFigFont{10}{12.0}{\rmdefault}{\mddefault}{\updefault}$v$}}}
\put(5918,-1673){\makebox(0,0)[b]{\smash{\SetFigFont{10}{12.0}{\rmdefault}{\mddefault}{\updefault}$v$}}}
\end{picture}

%% file: easy_open.pstex_t
\begin{picture}(0,0)%
\epsfig{file=easy_open.pstex}%
\end{picture}%
\setlength{\unitlength}{3947sp}%
\begingroup\makeatletter\ifx\SetFigFont\undefined%
\gdef\SetFigFont#1#2#3#4#5{%
  \reset@font\fontsize{#1}{#2pt}%
  \fontfamily{#3}\fontseries{#4}\fontshape{#5}%
  \selectfont}%
\fi\endgroup%
\begin{picture}(2574,1677)(5539,-2848)
\put(6826,-1336){\makebox(0,0)[b]{\smash{\SetFigFont{12}{14.4}{\rmdefault}{\mddefault}{\updefault}$v$}}}
\end{picture}